\documentclass{article}\usepackage[]{graphicx}\usepackage[]{xcolor}
\makeatletter
\def\maxwidth{ %
  \ifdim\Gin@nat@width>\linewidth
    \linewidth
  \else
    \Gin@nat@width
  \fi
}
\makeatother

\definecolor{fgcolor}{rgb}{0.345, 0.345, 0.345}

\usepackage{framed}
\makeatletter
 {\par\unskip\endMakeFramed%
 \at@end@of@kframe}
\makeatother

\definecolor{shadecolor}{rgb}{.97, .97, .97}
\definecolor{messagecolor}{rgb}{0, 0, 0}
\definecolor{warningcolor}{rgb}{1, 0, 1}
\definecolor{errorcolor}{rgb}{1, 0, 0}

\usepackage{alltt}
\usepackage{times}
\usepackage{moreverb}
\usepackage{amsmath}
\DeclareMathOperator{\sgn}{sgn}
\usepackage{mathabx}

\title {Robust regression techniques for multiple method comparison and transformation}

\author{Florian Dufey\\ 
\small{Roche Diagnostics GmbH, Assay Development \& System Integration (DSRIBF),}\\ \small{Nonnenwald 2, 82377 Penzberg / Germany}}
\date{\today}
\IfFileExists{upquote.sty}{\usepackage{upquote}}{}
\begin{document}
\maketitle
\begin{abstract}
    A generalization of Passing--Bablok regression is proposed for comparing multiple measurement methods simultaneously.
    Possible applications include assay migration studies or interlaboratory trials.
    When comparing only two methods, the method reduces to the usual Passing-Bablok estimator.   
    It is close in spirit to reduced major axis regression, which is, however, not robust. 
    To obtain a robust estimator, the major axis is replaced by the (hyper-)spherical median axis. 
    The method is shown to reduce to the usual Passing--Bablok estimator if only two methods are compared. 
    This technique has been applied to compare SARS-CoV-2 serological tests, bilirubin in neonates, and an \emph{in vitro} diagnostic test using different instruments, sample preparations, and reagent lots. 
    In addition, plots similar to the well-known Bland--Altman plots have been developed to represent the variance structure.
   \end{abstract} 
\maketitle
\footnotetext{\textbf{Abbreviations:} ARE: asymptotic relative efficiency, MAR: Major axis regression,  OLS: Ordinary least squares regression, PBR: Passing--Bablok regression, RMR: reduced major axis regression, mPBR, mMAR, mRMR: multivariate PBR, MAR and RMR, respectively, MC: Method comparison}

\section{Introduction}

In clinical chemistry, the comparison of different measurement procedures for the same analyte is a recurring task. 
In this context, the measurement procedures are often referred to as "methods" and corresponding studies are known as "method comparison (MC) studies" \cite{altman1983measurement, bland1986statistical}.
Typically, a set of patient samples is measured using both measurement procedures that are being compared, and some variant of linear regression is performed. 
Since both methods usually produce results with a comparable measurement error, 
the preferred analytical tools are regression methods commonly referred to as "errors in variables" models or "regression with measurement error" \cite{cheng1999statistical}.

Various review articles \cite{linnet1990estimation,linnet1993evaluation,francq2014measurement,bolfarine2020regression} and guidelines \cite{assay2013,clinical2013ep09} recommend specific regression methodologies for these studies. 
The recommended approaches are either variants of major axis regression (MAR), also known as Deming or orthogonal regression, or reduced major axis regression (RMR), also known as least product regression \cite{kummell1879reduction,adcock1878problem}. 
Alternatively, if a more robust regression method is required to handle outlying measurements or non-linear deviations near the boundaries, Passing--Bablok regression (PBR) is recommended \cite{passing1983new,bablok1988general,dufey2020derivation}. 
In PBR, the slope estimator is obtained as a shifted median of all pairwise slopes.

While there are multidimensional extensions of  MAR and RMR  \cite{Feldmann1981,bolfarine2020regression} (mMAR and mRMR, respectively),  which allow for the comparison of multiple measurement methods at the same time, there are currently no extensions of the  PBR to the multiple method comparison case. 
The aim of this article is to extend the robust PBR methodology to cases where measurements obtained from more than two methods need to be compared.
Additionally, the slope estimators to be obtained are required to be compatible in the sense that  if $\hat{\beta}_{12}$ is the slope estimate from comparing method 1 and method 2, and $\hat{\beta}_{23}$ is the slope estimate between method 2 and method 3, then we require that the estimator of the slope between method 1 and method 3, denoted as $\hat{\beta}_{13}$, fulfills the relationship $\hat{\beta}_{13}=\hat{\beta}_{12}\hat{\beta}_{23}$.
It is worth noting that while the MAR and RMR estimates are compatible in this sense, the pairwise PBR estimates are not.

To address this issue, we propose a multidimensional Passing-Bablok regression (mPBR) method, which reduces to the ordinary PBR in two dimensions and provides compatible and robust estimates in more than two dimensions. 
In mPBR, we utilize the (hyper-)spherical median axis instead of the major axis used in mRMR. 
The (hyper-)spherical median axis is a concept derived from directional statistics and is defined as an axis with minimal angular distance to all points after projecting the points and the median axis onto a (hyper-)sphere \cite{fisher1993,mardia2000directional}.

We will apply this method to the comparison of SARS-CoV-2 serological tests \cite{FERRARI2021144}, plasma bilirubin tests in neonates, and a clinical test that involves different sample preparation methods, instrument platforms, and reagent lots. 
In the latter example, we will demonstrate the utility of mPBR in analyzing differences between sub-groups.

\section{Mathematical derivation of the new regression method}

Consider a sample panel of size $n$ where individual samples are labeled by $i \in \{1,\ldots,n\}$  each of which is measured with $N$ different methods whose results are arranged into an array with components  $x_{i\mu}$ with $\mu \in \{1,\ldots,N\}$. 
The values $x_{i\mu}$ are assumed to follow the functional model:

\[
    x_{i\mu} = \beta_\mu r_i +\alpha_\mu+\epsilon_{i\mu},
\]

where $\beta:=\{\beta_\mu\}$ is the slope vector, the scalar $r_i$ is proportional to the unknown true concentration of sample $i$.
The values $r_i$ are assumed fixed constants, also called latent variables. 
$\alpha:=\{\alpha_\mu\}$ is the intercept vector, and $\epsilon_i:=\{\epsilon_{i\mu}\}$ is the vector of random errors.  
Note that not only $\alpha$ and $\beta$, but also the $r_i$ related to concentration, are parameters that need to be estimated. 
To make the parameters unique, one degree of freedom of $\beta$ and $\alpha$ may be fixed at will, e.g.\ $\beta_1=1$ and $\sum_\mu\alpha_\mu=0$.
Furthermore, the parameters $\beta_\mu$ and their estimates $\hat{\beta}_\mu$ are assumed to be positive. 
The random errors are assumed independent for $i\neq j$, while the distributions of $\epsilon_{i\mu}$ and $\epsilon_{i\nu}$ are similar and independent for $\mu \neq \nu$, so that the variance matrix $\Xi_i=\Gamma\sigma_{i\mu=1}^2$ (where $\Gamma$ is not to be confused with the Gamma function) is diagonal, and the variance ratios for different methods is independent of $i$. 

Otherwise the error distributions may be quite arbitrary, especially since
normality is not assumed and, due to the $\sigma_i:=\sigma_{i\mu=1}$ which may differ from sample element to element, the errors may be heteroscedastic. 
This generalizes the assumptions made in Bablok et al.\ \cite{bablok1988general} for the comparison of two methods. 

An useful equivalent expression is 
\[
    y_{i\mu}:=b_\mu x_{i\mu} = r_i +\epsilon'_{i\mu} + a_\mu,
\]
with the vector of inverse slopes $b :=\{1/\beta_\mu \}$, which may be interpreted as scaling factors, the scaled intercept parameters are $a_\mu:=b_\mu \alpha_\mu$ and $\epsilon'_{i\mu}:=b_\mu \epsilon_{i\mu}$.
The scaled values  $y_{i\mu}:= b_\mu x_{i\mu}$ are then symmetrically distributed in space around the axis $y(\lambda)=\lambda e+a$ along the space diagonal $e$ with  $e_\mu=1/\sqrt{N}$ for all $\mu$. 

\subsection{Derivation of mMAR and mRMR with zero intercept} 

First, let us consider the situation $a=0$, i.e., when the expected regression line passes through the origin. 
The estimated slope/ scaling vector $\Hat{\beta}$ is obtained from centered principal component analysis: 
The equations of mMAR, are obtained by minimizing the sum of squares  
\begin{equation}
    \label{eq:Deming}
     \sum_i(x_i-\beta r_i)^T \Xi_i^{-1} (x_i-\beta r_i)
\end{equation}
with respect to the $r_i$ and $\beta$. 
Here, $x^T$ signifies the transpose of $x$. 
This yields the estimates
\[
    \hat{r}_i=\frac{\hat{\beta}^T \Gamma^{-1} x_i}{\hat{\beta}^T\Gamma^{-1}\hat{\beta}}
    \]
and subsequently an equation for $\hat{\beta}$:  

\begin{equation}
    \label{eq:Demingbeta}
    \sum_i \sigma_i^{-2}(\hat{\beta}^T\Gamma^{-1}x_i)\left(I_N-\frac{\Gamma^{-1/2}\hat{\beta}\hat{\beta}^T\Gamma^{-1/2}}{\hat{\beta}^T\Gamma^{-1}\hat{\beta}}\right) \Gamma^{-1/2}x_i=0,
\end{equation}
where $I_N$ is the $N\times N$ diagonal unit matrix. 

This is just the well known equation of weighted mMAR with $s:=\Gamma^{-1/2}\hat{\beta}$ being the vector pointing in the direction of the major axis. 
Measurements with a large value of $r_i$ will have a high leverage, hence the method is not robust against outliers.

The equation for the mRMR slope estimate $\hat{\beta}$ follows with the special choice\footnote{A direct minimization of the sum of squares with this choice of $\Gamma$ and $\Xi$ would yield inconsistent results, cf.\ \cite{pronzato2013design}. }  $\Gamma_{\mu\mu}=\hat{\beta}^2_\mu$ in Eq.\ \ref{eq:Demingbeta}, i.e., $\Gamma^{-1/2}$ is a diagonal matrix   $\Gamma^{-1/2}=\mathrm{diag}(\hat{b})$ (we assume all $\hat{b}$ to be positive). 
Introducing the diagonal matrix $X_i=\mathrm{diag}(x_i)$ we have the identity 
\begin{equation}
    \label{eq:X_i}
    \Gamma^{1/2}x_i=X_i\hat{b}.
\end{equation}
Finally,
\begin{align}
    \label{eq:beta}
    \sum_i \sigma_i^{-2}(e^T\Gamma^{1/2}\hat{x}_i)\left(I_N-ee^T\right) \Gamma^{1/2}\hat{x}_i&=0\text{, or,}\\
    \sum_i \sigma_i^{-2}(e^TX_i\hat{b})\left(I_N-ee^T\right) X_i\hat{b}&=0.
\end{align}

\subsection{Derivation of mPBR with zero intercept} 

\begin{figure}[ht]
    \begin{centering}
        \includegraphics[width=0.8\textwidth]{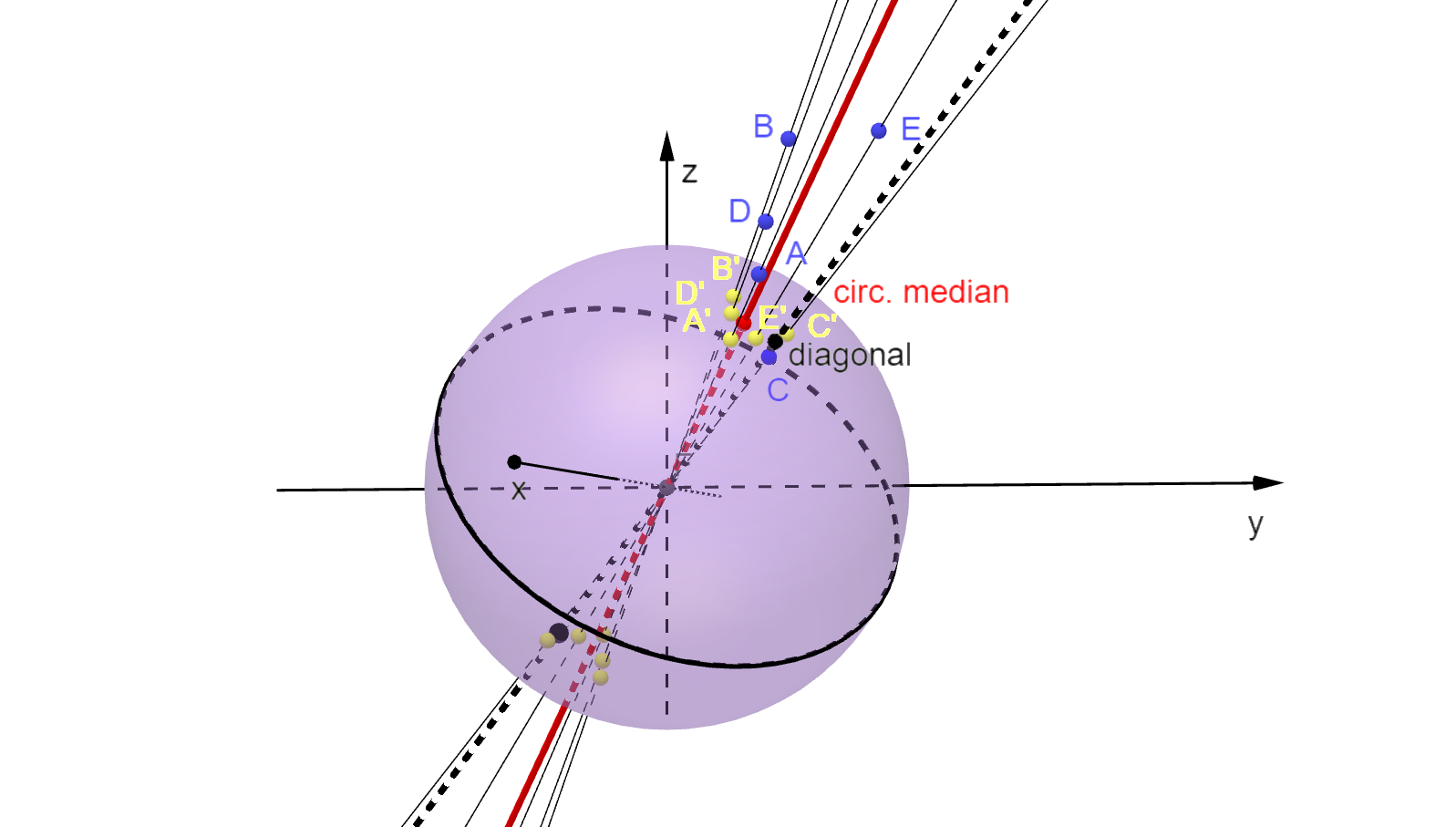}
        \caption{\label{fig:Spherical}Spherical median axis (red) of 5 points (A--E, blue). 
        The yellow points A'--E' are the projections of the points on the sphere. 
        The red point corresponding to $s$ designates the position of the spherical median. 
        It has the smallest angular distance on the sphere to the points A'-- E'. 
        Only the points on the same hemisphere as $s$ are taken into account. 
        $\Gamma=I$ was assumed for illustration.  }
    \end{centering}
\end{figure}

The alternative mPBR algorithm utilizes ideas from finding the spherical median axis \cite{fisher1993}, cf.\ Fig.\ \ref{fig:Spherical}. 
In the spirit of most robust regression methods, all weights are assumed equal $\sigma_i^{-2}=1$. 
Consider first the situation analogous to mMAR, where the matrix $\Gamma^{-1}$ is assumed to be diagonal, its element are known and do not need to be estimated.  

We consider straight lines passing through the origin and the scaled points $\Gamma^{-1/2}x_i$. 
With $\tilde{x}:= x/|x|$ signifying a vector of unit length with the same direction a $x$ (not to be confused with the median),
each of these lines will intersect the unit sphere at two antipodal points $\pm\widetilde{\Gamma^{-1/2}x}_{i}= \pm\Gamma^{-1/2} x_i/\lvert\Gamma^{-1/2} x_i\rvert $. 
The circular median axis is then the line defined by the unit vector $\tilde{s}$ ( remember $s=\Gamma^{-1/2}\hat{\beta}$) with smallest sum of distances to all points $\widetilde{\Gamma^{-1/2}x}_i$ on the same hemisphere.
Mathematically, we seek to minimize the following expression: 
\begin{equation}
    \label{eq:arccos}
    \underset{\tilde{s}}{\arg\min}\sum_i\arccos\left(\left\lvert \tilde{s}^T\widetilde{\Gamma^{-1/2}x}_i\right\rvert\right).
\end{equation}
For the case where $\Gamma$ is a unit matrix, the situation is illustrated in Fig.\ \ref{fig:Spherical}.

We might call this figuratively the sky-mongers regression: 

Thinking of the points $\widetilde{\Gamma^{-1/2}x}_i$ as positions of the stars of some constellation on the celestial sphere, and the point $\tilde{s}$ on the sphere as the center of the respective constellation, fixing the line of sight of the sky-monger (in the center of the celestial sphere) to $\pm\tilde{s}$ on the sphere.
However, unlike  traditional astronomical constellations, the projections of the points for which a regression axis is sought  may well be bipolar, similar finding a line with minimal angular distance to both the Big Dipper on the northern hemisphere and the Southern Cross on the southern hemisphere simultaneously. 

As long as none of the $\widetilde{\Gamma^{-1/2}x}_i$ falls directly on the axis defined by $\tilde{s}$, the dependence on $\tilde{s}$ is analytical, and the minimum is attained if  
\begin{equation}
    \label{eq:optimum}
    \sum_i \sgn(\tilde{s}^T \Gamma^{-1/2}x_i)\frac{(I_N-\tilde{s}\tilde{s}^T) \Gamma^{-1/2}x_i}{\left\lvert (I_N-\tilde{s}\tilde{s}^T) \Gamma^{-1/2}x_i\right\rvert}=0.
\end{equation}

While the corresponding regression estimate $\hat{\beta}\propto \Gamma^{1/2}\tilde{s}$ is robust against outlying measurements, it will be an unbiased estimate of $\beta$ only if the errors follow a normal distribution. 
Therefore, the equivalent of the mRMR, obtained by setting $\Gamma_{\mu\mu}=\hat{\beta}^2_\mu$ in Eq.\ \ref{eq:optimum} is more attractive. 
With this choice, the scaled errors $\epsilon'_{i\mu}$ are all from the same distribution for each $i$, so that the distribution of the $y_{i\mu}-a_\mu$ is symmetrical around the $r_i$.
This seems sufficient to guarantee consistency of the mPBR estimator $\hat{b}$.   

In this case, the spherical median  $\tilde{s}$ of the $y_i$ after scaling,  is known to coincide with the space diagonal $e$,
so the Eq.\ \ref{eq:optimum}  becomes an equation for the estimated scaling vector $\hat{b}$:

\begin{equation}
    \label{eq:arccos2}
    \underset{\tilde{s}}{\arg\min}\sum_i\arccos\left(\left\lvert \tilde{s}^T\widetilde{X_i \hat{b}}\right\rvert\right)=e,
\end{equation}
where $X_i$ was introduced in Eq.\ \ref{eq:X_i}. 

Hence, The equation for estimating $\hat{b}$ becomes: 
\begin{equation}
    \label{eq:withoutinter}
    \sum_i\sgn(x_{i}^T\Hat{b}) \frac{PX_i \hat{b}}{\left\lvert PX_i\hat{b}\right\rvert}=0,
\end{equation}
where $P=(I_N-ee^T)$ is the projection matrix to the hyperplane perpendicular to $e$.
Comparing Eqs.\  \ref{eq:beta} and \ref{eq:withoutinter}, we can see that the latter differs from the former by the missing weighting with $\sigma_i$, the replacement of $e^TPX_i\hat{b}$ by its sign, and the replacement of $PX_i\hat{b}$ by a unit vector pointing in the same direction. 
The influence of each term is bounded by $\pm 1$, making the estimate robust. 

\begin{figure}[ht]

{\centering \includegraphics[width=0.8\textwidth]{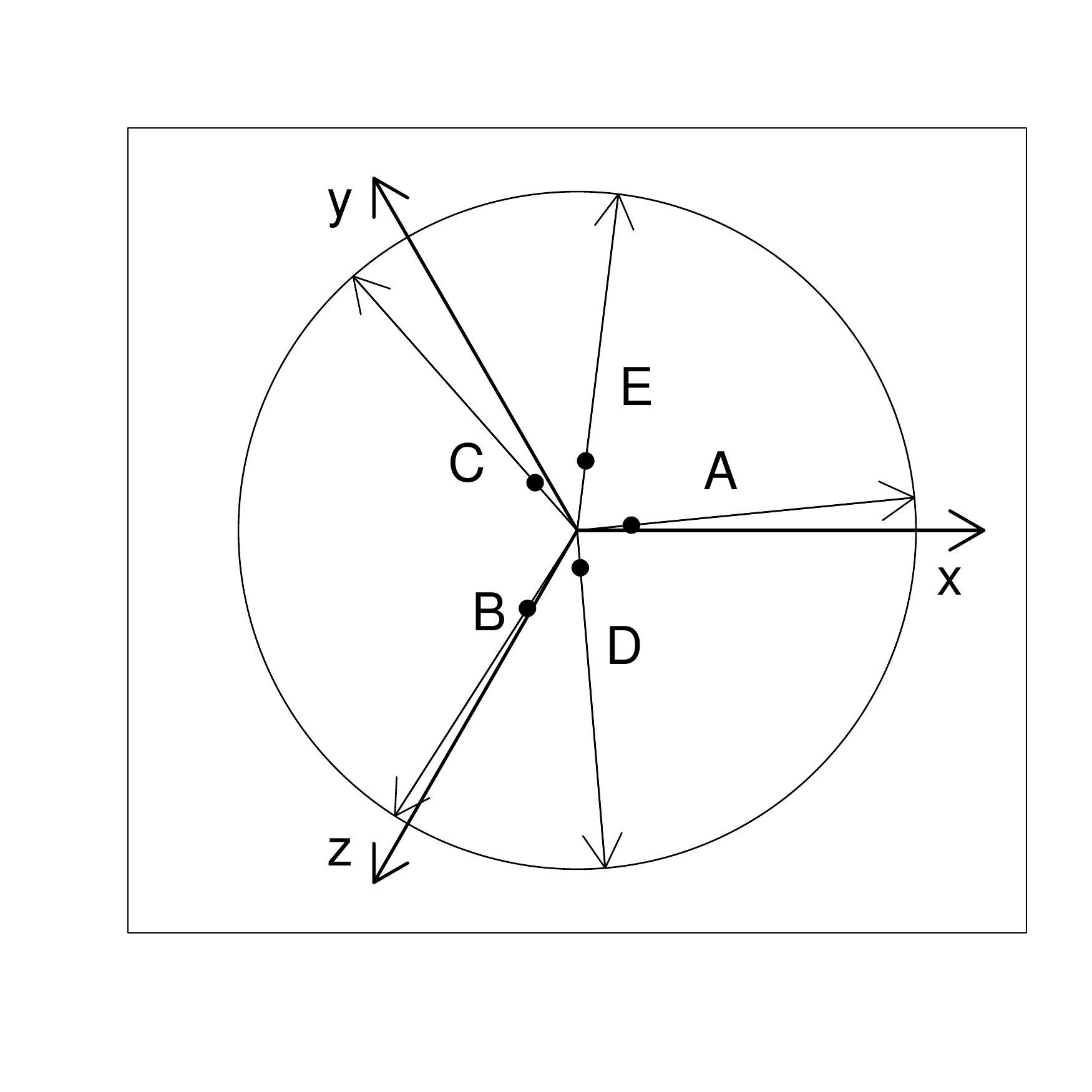} 

}

\caption[Visualisation of the mPBR algorithm with zero intercepts]{Visualisation of the mPBR algorithm with zero intercepts: The points are scaled with the $\hat{b}_\mu$ until the unit vectors passing through the points $A-D$ (filled dots) in the plane perpendicular to the space diagonal sum up to 0.}\label{fig:SphericalFig}
\end{figure}

The actual slope estimator of  method $\nu$ versus method $\mu$ is $\Hat{\beta}_{\mu\nu}=\Hat{b}_\mu/\Hat{b}_\nu$. 

These estimators fulfill the compatibility condition $\Hat{\beta}_{\mu\nu}\Hat{\beta}_{\nu\kappa}=\Hat{\beta}_{\mu\kappa}$. 
 
By choosing $\Hat{b}_{1}=1$, the scale of $b$ and $r$ is set to that of the first method. 
Then, the $\Hat{\beta}_\mu=1/\Hat{b}_\mu$ with $\mu >1$ represent the $N-1$ slopes relative to that of the first method with $\Hat{\beta}_{1}=1$.

\subsection{Derivation of mPBR with non-zero intercept}

If the intercept parameters are a priori not known to be 0, then,  in the spirit of Theil--Sen and Passing--Bablok regression, the estimation of the scaling parameters $b$ can be decoupled from the estimation of the intercept parameters $a$.
To this end, we replace the $x_i$ in Eq.~\ref{eq:withoutinter} with the differences between the x-values, denoted as  $\Delta x_J:= x_j-x_i$ with $i<j$. The sum is extended to the corresponding index pairs $\{J\}:=\{(i,j)\}$. 
As before (Eq.\ \ref{eq:X_i}), diagonal matrices $\Delta X_J$ are defined with $\Delta X_{J}=\mathrm{diag}(\Delta x_{J})$. 
We get the final equation determining the slope vector $\hat{b}$ of the mPBR method: 

\begin{equation}
    \label{eq:withinter}
    \sum_J\sgn(\Delta X_J^T\Hat{b}) \frac{P\Delta X_J \hat{b}}{\left\lvert P\Delta X_J\hat{b}\right\rvert}=0.
\end{equation}

In two dimensions, with $x'_i:=x_{i\mu=1}$, $y'_i:=x_{i\mu=2}$, $\Hat{\beta}_1=1$ and $\Hat{\beta}_2=\hat{\beta}_{12}$,
Eq.\ \ref{eq:withinter} simplifies to: 
\[
    \sum_J\sgn(\Delta y'_J+\hat{\beta}_{12}\Delta x'_J) \sgn(\Delta y'_{J} -\Hat{\beta}_{12} \Delta x'_J))=0.
\]

This equation coincides with the defining equation of the pairwise PBR \cite{dufey2020derivation}.

With the choice  $\sum_\mu a_\mu= e^Ta =0$ made for the intercept vector, 
\[
    \mathrm{E}\left( \frac{x_{i}^T b}{N}\right)=r_i
\]
we use
\[ 
\Hat{r}_i = \frac{x_{i}^T\Hat{b}}{N}
\]
as an estimator for $r_i$. 
Then we obtain a robust estimator for  the intercept vector $\Hat{a}_\mu$ from the spatial median, denoted as $\mathrm{SpMed}$ \cite{weiszfeld1937},  
\[
    \Hat{a}_\mu= \mathrm{SpMed}\left(\{Px_i \Hat{b}\} \right).
\]

The spatial median $\mathrm{SpMed}$ is defined as the point with minimal distance from all the $x_i$: 
\[
    \label{eq:SpMed}
    \mathrm{SpMed}(\{x_i\}) :=\underset{y}{\arg\min} \sum_i d(x_i-y),
\]
where $d(x)$ is the usual Euclidean norm of the vector $x$. 
This is equivalent to the equation:  
\[
    \label{eq:SpMed0}
    \sum_i \frac{x_i-\Hat{y}}{|x_i-\Hat{y}|}=0, 
\]
as long as $\hat{y}$ does not equal any of the $x_i$ (adaptations for this situation are easily devised).

From the intercept parameter vector $\hat{a}$ the intercept $\alpha_{\mu\nu}$ for the graph of $x_\nu$ vs.\ $x_\mu$, $\hat{\alpha}_{\mu\nu}$, can be calculated as
\[
    \hat{\alpha}_{\mu\nu}:=\frac{\hat{a}_\nu-\hat{a}_\mu}{\hat{b}_\nu}=\hat{\alpha}_\nu-\hat{\alpha}_\mu \frac{\hat{b}_\mu}{\hat{b}_\nu}.
\]

\subsection{Computation of the mPBR estimates}

A solution of Eq. \ref{eq:withinter} (and similarly Eq.\ \ref{eq:withoutinter}) can be achieved through iterative reweighting \cite{weiszfeld1937,beck2015weiszfeld}. 
To solve Eq.\ \ref{eq:withinter}, 
approximate values for the weights 
\[
    w_J^{(n)} =  \frac{\sgn(\Delta x_J^T\Hat{b}^{(n)})}{\left\lvert P\Delta x_J\hat{b}^{(n)}\right\rvert}
\]
can be calculated using the current estimator of the slopes $\Hat{b}^{(n)}$ (a starting value could be all $\Hat{b}^{(0)}_\mu=1)$.
Subsequently, an improved estimator  $\Hat{b}^{(n+1)}_\mu$ can be derived by solving the linear equation system:
\[
    \sum_J w_J^{(n)} P\Delta x_J \hat{b}^{(n+1)}=0.
\]

Similarly, the spatial median estimate of the intercept $\hat{a}$ can be computed.
This can be done either after obtaining a converged estimate of the slope from 
\[ 
    \sum_i v_i^{(n)} P (\hat{a}^{(n+1)}-x_i \hat{b})=0
\] 
with weights $v_i^{(n)}=|P (\hat{a}^{(n)}-x_i \hat{b})|$, 
or simultaneously by replacing $\hat{b}$ with $\hat{b}^{(n+1)}$ in these equations. 
It has been proven that slight extensions of the Weiszfeld algorithm \cite{beck2015weiszfeld}  for the spatial median always converge. 
Eventually, it is possible to implement similar extensions for slope estimation. 
The simultaneous procedure offers the advantage of enabling to impute missing observations in the spirit of an expectation-maximization (EM) algorithm \cite{dempster1977maximum}.

\subsection{Statistical properties of the multivariate PBR estimator }
\label{sec:StatProp}
The examples presented in the next section showcase the potential of the new method. 
However, a formal proof of its statistical properties is not in scope of this article. 
Nevertheless, we at least want to discuss some concepts and compare to related estimators.

 \begin{description}
     \item[Consistency] We can assume that the mPBR estimator is consistent due to the symmetry of the model with respect to exchange of the measurements. 
         Namely, for each $i$ all $y_{i\mu}$ are distributed identically.  

     \item[Finite sample bias] It is known \cite{dufey2020derivation} that the estimates for the components of $b$ are biased in case of RMR and also ordinary PBR, while the logarithms $\ln \hat{b}$ are unbiased. 
        It seems plausible that this also the case for the estimators $\Hat{b}$ from mPBR in more than two dimensions. 

    \item[Asymptotic efficiency]  Fisher \cite{fisher1993} reports an  asymptotic relative efficiency (ARE) of $\pi/4$   for the spherical median axis assuming the distribution of the points on the sphere to follow a Watson distribution in the limit of small dispersion parameter.
        This coincides with the ARE of the spatial and spherical median.
         Comparing the variances of the PBR to that of the RMR in two dimensions and assuming a Gaussian distribution, a relative asymptotic efficiency of $9/\pi^2$ was reported \cite{dufey2020derivation}. 

    \item[Robustness]  The mPBR slope estimator is robust against outlying points, as the angle on the sphere is bounded. 
        We expect the breakdown point not to depend on dimension, as intuitively, the worst situation occurs if all outliers pull in the same direction, which already happens in the two-dimensional case.  
        For the two-dimensional case, the breakdown point is $1-\sqrt{2}/2$, as for Theil--Sen regression \cite{dufey2020derivation}.

    \item[Numerical complexity] The proposed algorithm is not optimal in terms of  numerical complexity, as it scales quadratically with the number of measurements $n$. 
        However, in the two-dimensional case, a reduction to  quasilinear scaling with $n$ is possible \cite{raymaekers2022equivariant}. 
        Whether similar algorithms can be developed in the multidimensional case, is an interesting question. 

    \item[Uncertainty] The concept of  confidence limits for the slope in MC is highly problematic due to the Gleason phenomenon \cite{cheng1999statistical}. 
     Expressions for the asymptotic variance of the slope and intercept of the  PBR have been derived \cite{dufey2020derivation}.   
        Bootstrapping seems to be a valuable alternative \cite{raymaekers2022equivariant}. 
 \item[Identifiability] MC and errors in variable models face the challenge that the individual variances of the methods being compared are inestimable from the data used in regression \cite{reiersol1950identifiability}. 
         Even in situations where an estimation is theoretically possible, the corresponding estimators are based on higher-order moments, that are difficult to estimate with sufficient precision in practice.
         Therefore, most methods used for MC make assumptions about the error structure and variance ratios, and mPBR is no exception. 
         It is important to keep in mind that the smaller the correlation of the data, the larger the possible error due to misspecification of the variance ratio.
\end{description}

\section{Examples }
\label{sec:Examples}

\subsection{Multidimensional regression of SARS-CoV-2 measurement data }
\label{sec:SARS}

\begin{figure}[ht]

{\centering \includegraphics[width=0.8\textwidth]{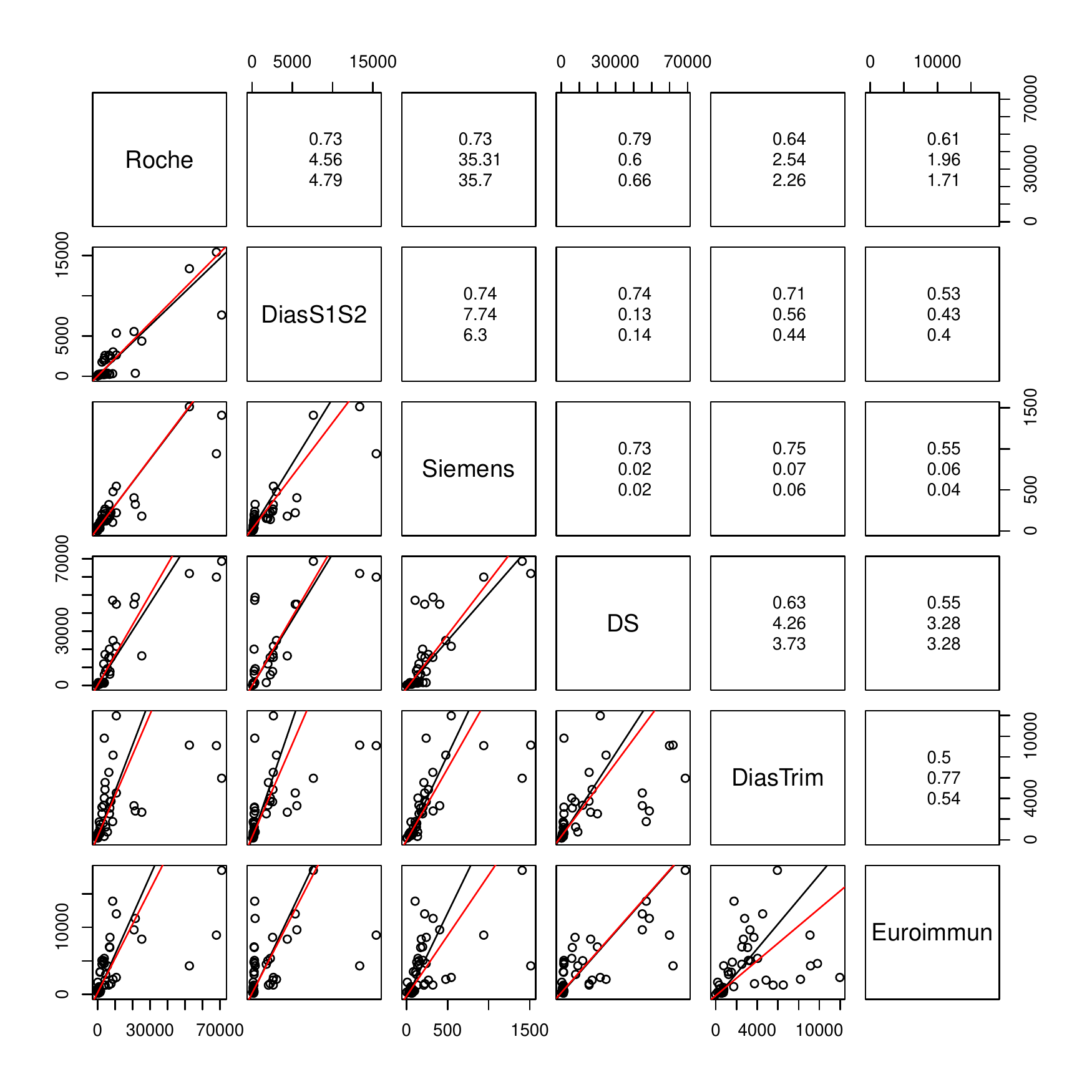} 

}

\caption[Comparison of mPBR (red lines) to pairwise PBR (black lines) in the SARS-CoV-2 test example]{Comparison of mPBR (red lines) to pairwise PBR (black lines) in the SARS-CoV-2 test example. In the upper triangle, Kendall's correlation coefficient $\tau$,  slope estimates obtained with mPBR (center) and pairwise PBR (bottom) are reported. The unit on all axes is U/mL. }\label{fig:SARS}
\end{figure}

The methodology developed above was applied to find a regression line through data (cf.\  Ferrari et al.\ \cite{FERRARI2021144}) from 48 samples measured with 6 different quantitative SARS-CoV-2 serological tests\footnote[2]{Full names of the tests and abbreviations used in the text and captions: Diasorin LIAISON SARS-CoV-2 Trimerics IgG (DiasTrim),  Euroimmun Anti-SARS-CoV-2  (Euroimmun), Roche Elecsys\textregistered Anti-SARS-CoV-2-S  (Roche), Diesse CHORUS SARS-CoV-2 ``NEUTRALIZING'' Ab (DS), Atellica IM SARS-CoV-2 IgG (Siemens), Diasorin LIAISON SARS-CoV-2 S1/S2 IgG (DiaS1S2)}  (Fig.\ \ref{fig:SARS}).
The estimates for the slopes obtained with the mPBR were compared to the pairwise Passing--Bablok estimates. 

The correlation between the different methods is generally quite poor, and even the assumption of a linear relation between some of the measurement pairs may be doubtful. Therefore, linear regression may not be the best way to analyze these data. 
Despite, or maybe because of these shortcomings, this data set seems well suited to visualize the differences between the pairwise and multivariate PBR.
Moreover, the incompatibility of the pairwise estimates can be considerable, contrasting with the compatibility of the mPBR estimates. 

\begin{figure}[ht]

{\centering \includegraphics[width=0.8\textwidth]{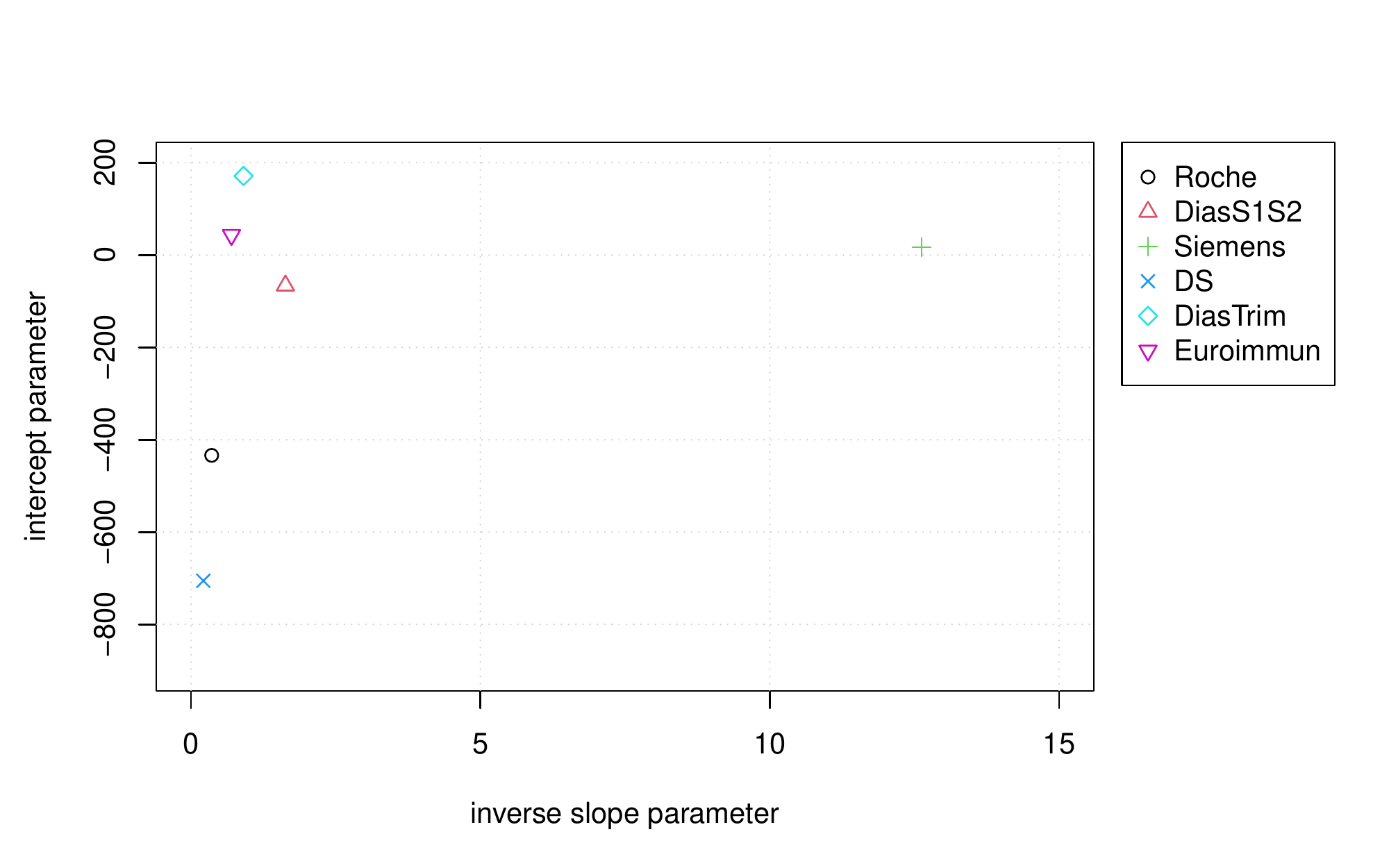} 

}

\caption[Intercept vs]{Intercept vs. inverse slope parameters of the various SARS-CoV-2 serological tests (refering to U/mL). Large differences, especially in $b$, can be explained by units U being manufacturer specific.}\label{fig:slopesvsintercept}
\end{figure}

Kendall's correlation coefficient $\tau$ ranges from 0.5 between the DiasTrim and the Euroimmun test and 0.79 between the Roche  and the  DS test.
Slope estimators obtained with the multivariate method are mostly less extreme, i.e. closer to 1 than the pairwise PBR estimates. 
While the mPBR estimators are compatible by design, there are considerable incompatibilities between the pairwise PBR estimators. 
For example, the estimators of DS vs.\ DiasTrim and Euroimmun are 3.73 and 3.28, respectively, from which an DiasTrim  vs.\ Euroimmun estimator of  $3.28/ 3.73=0.88$ can be calculated, which is more than 60\% larger than the one from direct comparison, which is 0.54. 
In Fig.\ \ref{fig:slopesvsintercept} the estimated intercept parameters are plotted versus the inverse slope parameters, providing a more condensed representation of the regression estimates. 
The large differences in slope between some of the assays are due to the use of assay-specific standards (U/mL), which are different from the international standard established by the WHO, who defined binding antibody units per milliliter (BAU/mL).
Therefore, this analysis is rather of the method transformation than the MC type and may be used co calculate conversion factors. 
For the Roche, Euroimmun and DS test, conversion factors of 1 BAU/U were reported by the manufacturers. 
For the Siemens and Diastrim  assay conversion factors of 21.8 BAU/U and 2.6 BAU/U, respectively, were reported. 
No conversion factor was reported for the DiaS1S2 test. 
If we neglect the relatively small intercepts, a conversion factor may be estimated for the DIAS1S2 test: 
Dividing the $\hat{b}_{\mu}$ values by the manufacturer-supplied conversion factors for the 5 tests for which the factor was available, and taking the geometric mean, yields a mean value $\bar{b}=0.404$ on the WHO BAU/mL scale. 
Dividing the DiaS1S2 inverse slope parameter  $\hat{b}=1.63$ by this mean yields a conversion factor of $4.04$ BAU/U for this test.

\begin{figure}[ht]

{\centering \includegraphics[width=0.8\textwidth]{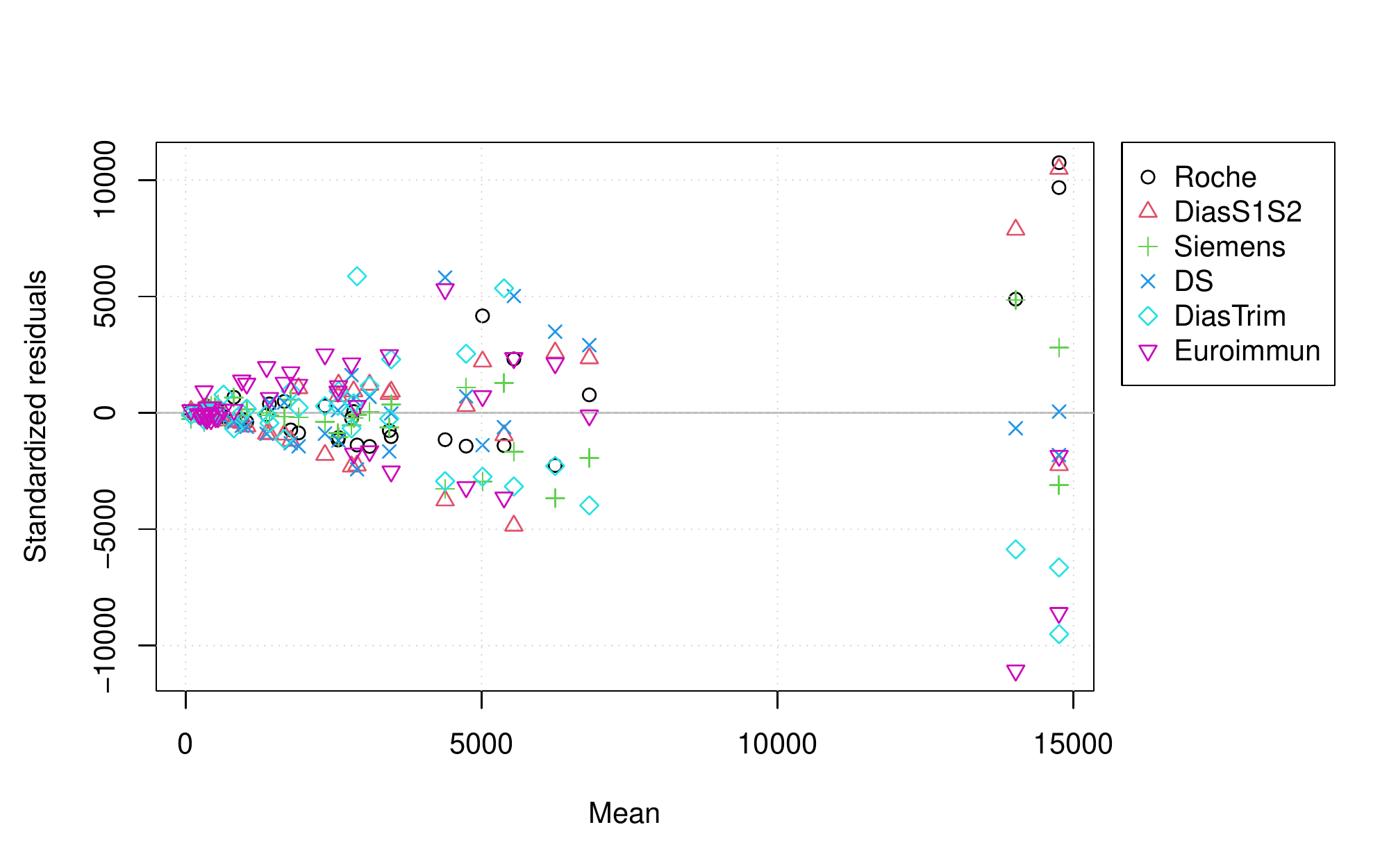} 

}

\caption[Generalized Bland--Altman plot for the SARS-CoV-2 serological tests with  $y_i-r_i-a_i$ (Standardized residuals) on the ordinate and  $r_i$ values (Mean) on the abscissa]{Generalized Bland--Altman plot for the SARS-CoV-2 serological tests with  $y_i-r_i-a_i$ (Standardized residuals) on the ordinate and  $r_i$ values (Mean) on the abscissa.}\label{fig:BlandAltmanMult}
\end{figure}

Finally, in Fig.\ \ref{fig:BlandAltmanMult}, the deviation (``Standardized residuals'') of the $y_i$ from the mean $\hat{r}_i+\hat{a}$, on a common scale, are plotted as a function of $\hat{r}_i$ (``Mean''), which is similar in spirit to Bland--Altman plots \cite{bland1986statistical}.   
The standard deviations seem to increase linearly with $r$. 
For mean values below 10,000, the errors seem to be highest for the Diastrim, DS and Euroimmun tests, with the DS and Siemens test showing some systematic bias.
At higher values ($r>10,000$), the Siemens and DS test show little bias, while the Diastrim and Euroimmun test appear biased to lower values, and the Roche and DiasS1S2 tests appear biased to higher values.

\subsection{Bilirubin in neonates} 
\label{sec:Neonates}

Warner et al.\ \cite{thomas2022total} conducted a study comparing total bilirubin measurements in plasma samples from neonates using assays and analysis platforms from different manufacturers\footnote[3]{Names of the analysis platforms and manufacturers: Alinity c and Architect ci16200 (Abbott Diagnostics, USA), AU5822 (Beckman Coulter, USA), cobas c702 and cobas c502 (Roche Diagnostics, Germany), ABL90 FLEX (Radiometer Denmark)}.
This article should also be consulted for further details on the assays and the data used below.

\begin{figure}[ht]

{\centering \includegraphics[width=0.8\textwidth]{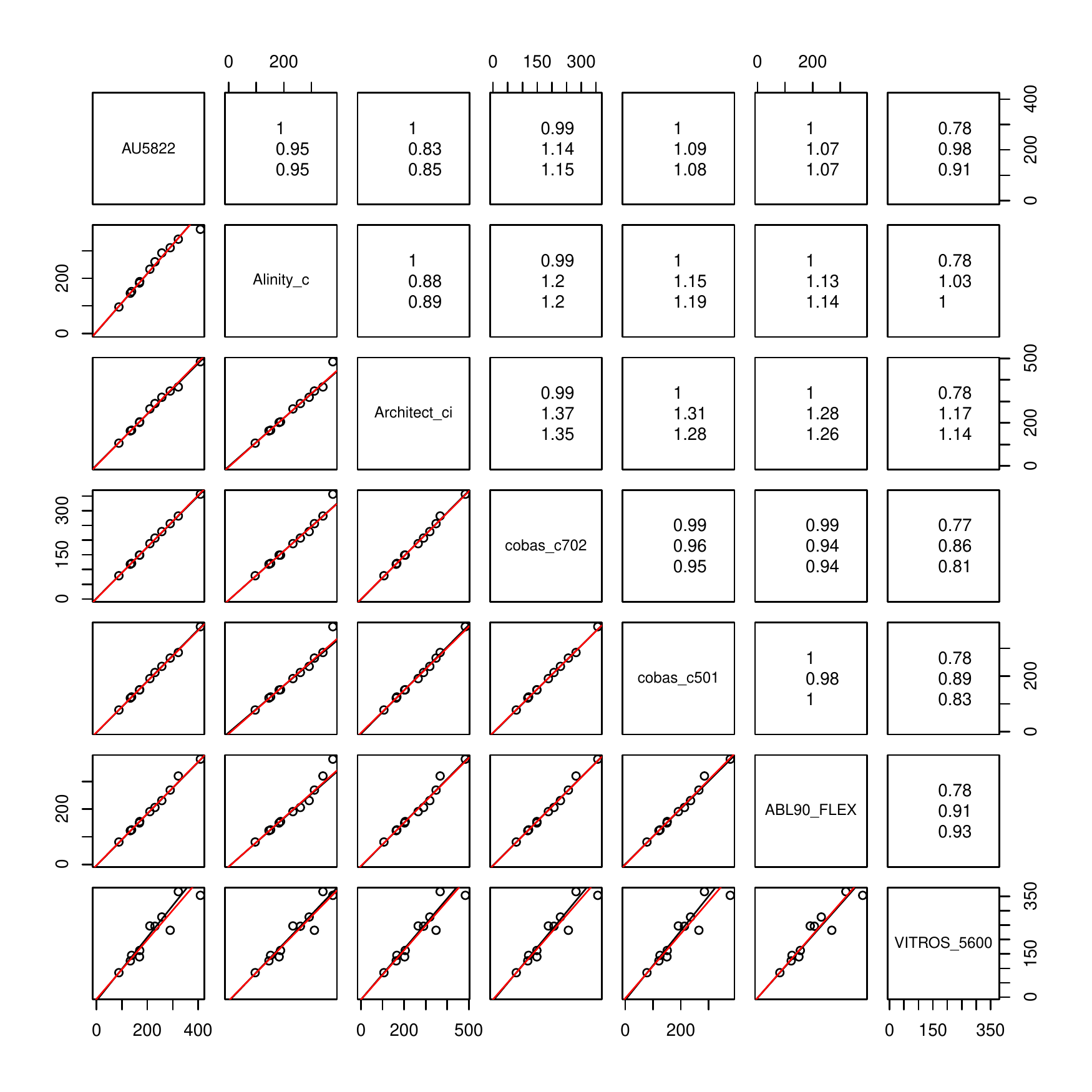} 

}

\caption[Comparison of 11 neonate total bilirubin plasma samples (concentration in $\mu$mol/L) with 7 different assays with mPBR (red lines) and pairwise PBR (black lines)]{Comparison of 11 neonate total bilirubin plasma samples (concentration in $\mu$mol/L) with 7 different assays with mPBR (red lines) and pairwise PBR (black lines). In the upper triangle, Kendall's correlation coefficient $\tau$,  slope estimates obtained with mPBR (center) and pairwise PBR (bottom) are reported.}\label{fig:neonates}
\end{figure}

While the precision within each method was high, the slopes of the PBR lines differed considerably from 1 for some methods,  which could potentially put neonates with high bilirubin plasma levels at risk of not receiving adequate therapy. 
To address this issue, the data were re-analyzed using mPBR. 
A total of 11 samples were measured, most of them in duplicates. 
However, as the second measurement was not available for all systems, only the first measurement was used in the calculations presented below.
The mPBR line parameters almost coincide with the ones obtained using pairwise PBR, with only slight differences visible for the comparisons to the VITROS 5600 system, as shown in Fig.\ \ref{fig:neonates}.

\begin{figure}[ht]

{\centering \includegraphics[width=0.8\textwidth]{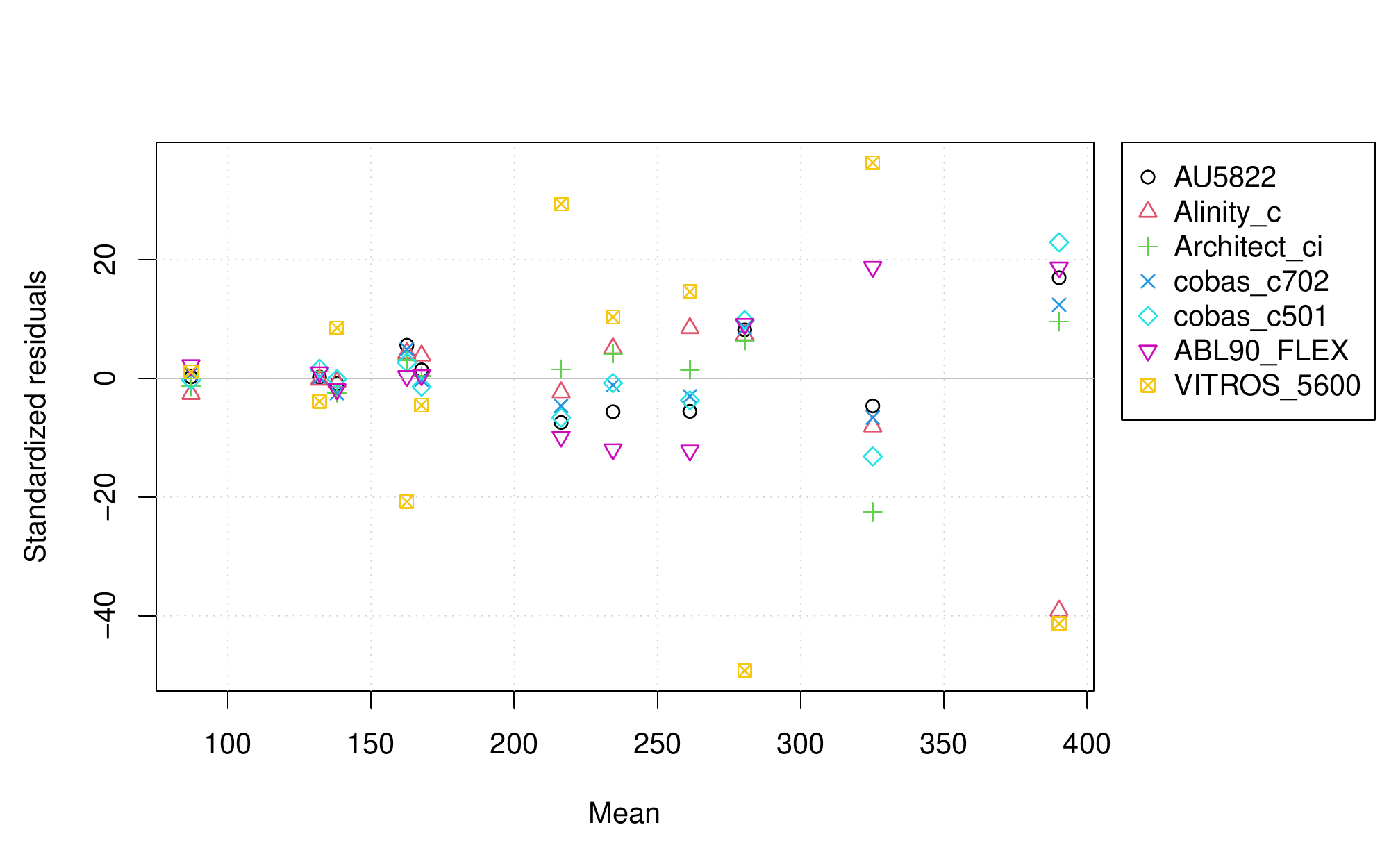} 

}

\caption[Residuals after scaling vs]{Residuals after scaling vs. mean  for the 11 bilirubin plasma samples with 7 different assays.}\label{fig:Neonateba}
\end{figure}

\begin{figure}[ht]

{\centering \includegraphics[width=0.8\textwidth]{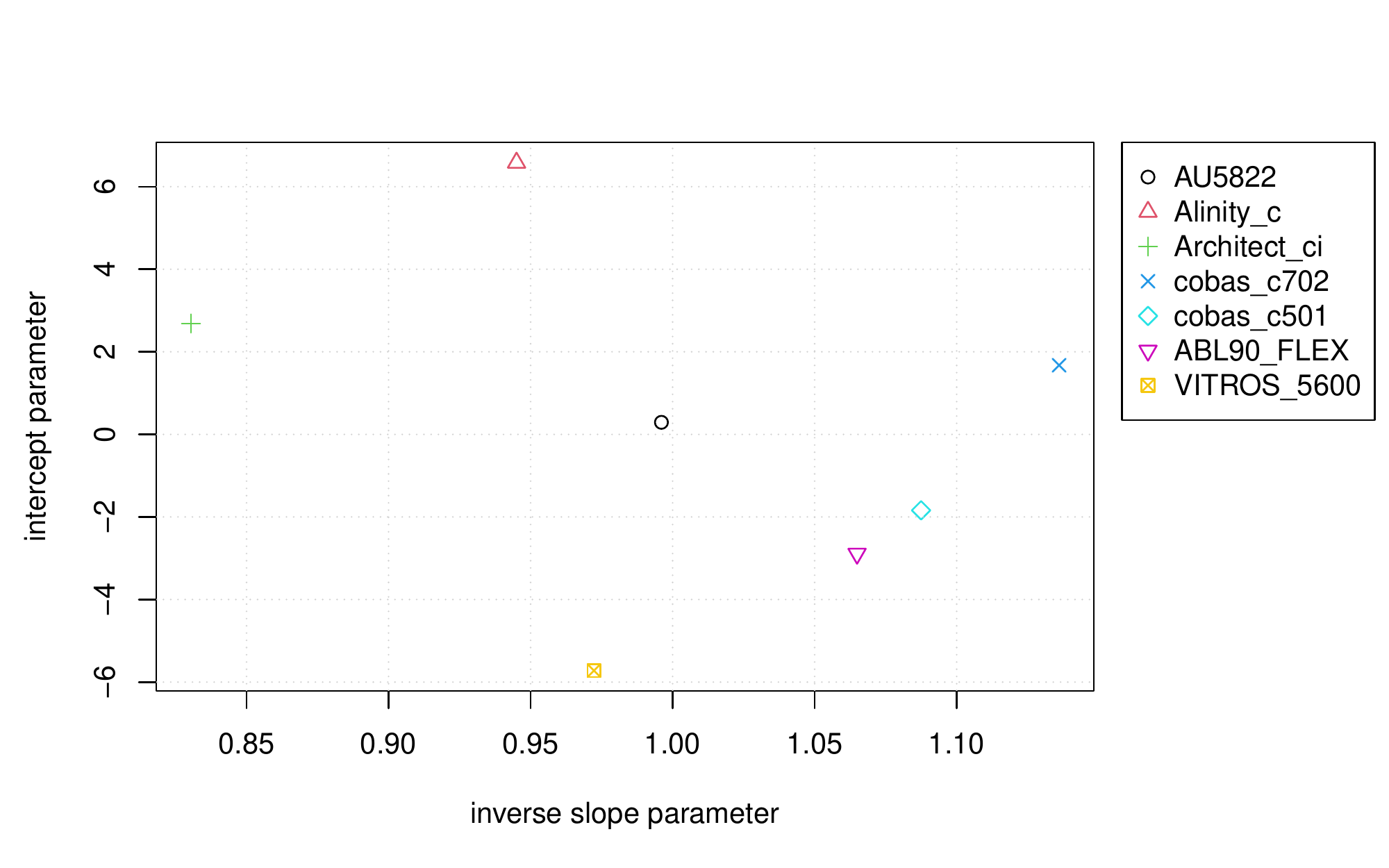} 

}

\caption[Intercept and inverse slope parameters of the 11 bilirubin plasma samples with 7 different assays]{Intercept and inverse slope parameters of the 11 bilirubin plasma samples with 7 different assays.}\label{fig:Neonatepara}
\end{figure}

Additionally, the VITROS 5600 system exhibits larger imprecision compared to the other systems, cf.\ Fig.\ \ref{fig:Neonateba}, confirming the findings in \cite{thomas2022total}. 
Similar to the previous example, mPBR allows for a condensed representation of the analysis results, as depicted in Fig.\ \ref{fig:Neonatepara}. 
The figure shows aconsiderable difference in inverse slope parameters, with the Architect ci 16200 on the lower end and the cobas\textregistered c702 on the other end.

\subsection{Comparison of an assay from clinical chemistry}
 \label{sec:Clinchem}

As a third example, the results of a multidimensional MC, involving two different sample preparation methods (A and B), two different measurement platforms (X and Y), and 6 different reagent lots (1 to 6) are presented for a quantitative clinical test using 150 whole blood samples. 
A Bland--Altman type plot of the residuals versus the mean  did not show any relevant differences between the assays, so it is not shown.

\begin{figure}[ht]

{\centering \includegraphics[width=0.8\textwidth]{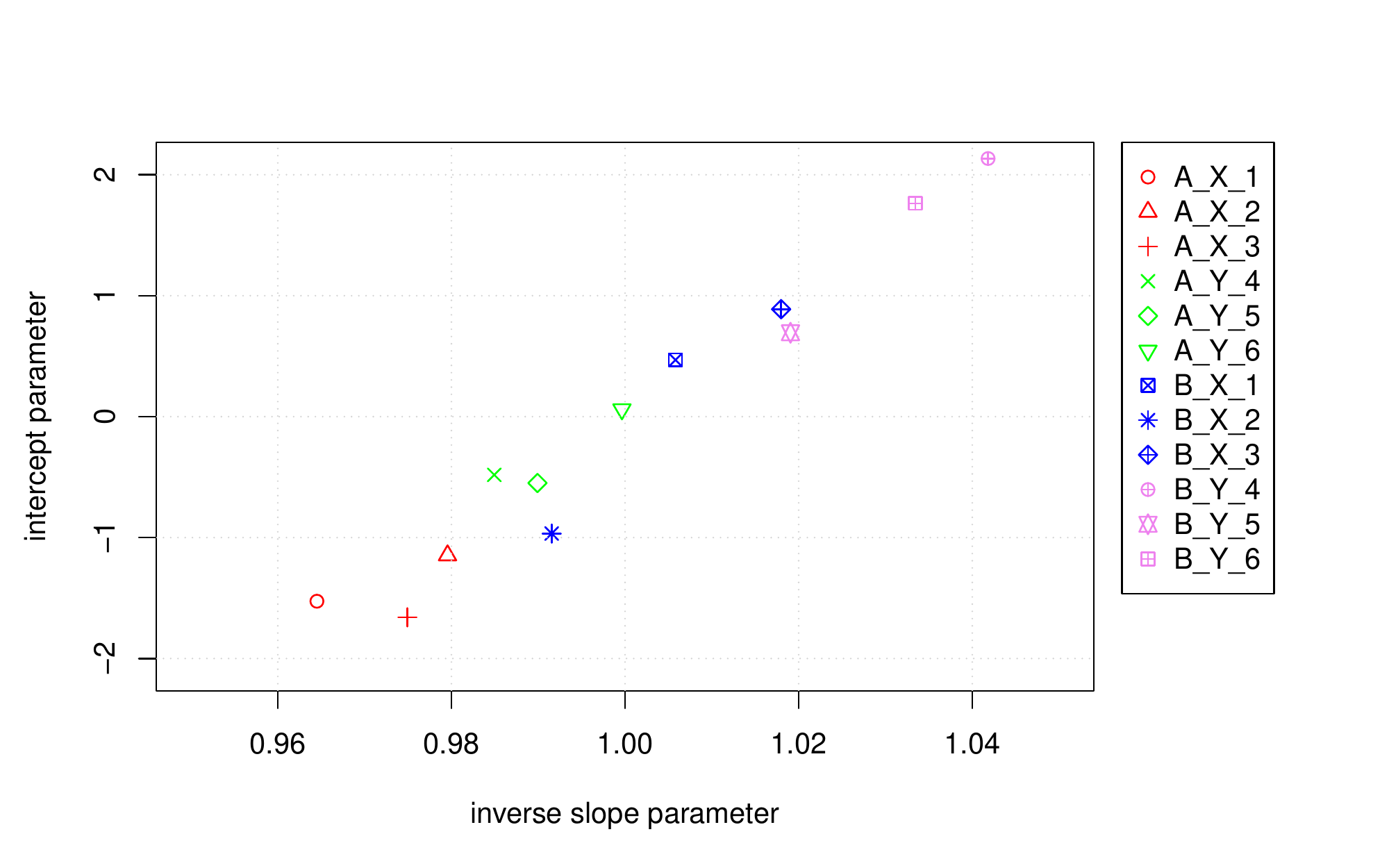} 

}

\caption[Intercept and inverse slope parameters  for the clinical assays differing in sample preparation (A vs]{Intercept and inverse slope parameters  for the clinical assays differing in sample preparation (A vs. B), instrument platform (X vs. Y), and reagent lot (1 to 6). Systematic differences between both instrument platform and sample preparation are clearly visible.}\label{fig:Clinpara}
\end{figure}

However, differences between the two sample preparation methods and, to a lesser extent, between the instrument platforms are clearly visible in the plot of the intercept vs.\ inverse slope parameters, as depicted in Fig.\ \ref{fig:Clinpara}.

To quantify these differences further, an OLS regression analysis can be performed using the inverse slope parameters and intercept parameters as the response variables, and the relevant effects (pre-analytical treatment and instrument platform) as the explanatory variables.
Let $x_\mathrm{pre}$ be the dichotomic variable representing the pre-analytical treatment, with levels 0 for comparator A and 1 for B. Similarly, let $x_\mathrm{inst}$ be the dichotomic variable representing the instrument platform, with levels 0 for comparator X and 1 for Y. The linear model

\[
    -\mathrm{ln}(b) \sim q_\mathrm{pre}x_\mathrm{pre}+q_\mathrm{inst}x_\mathrm{inst}
\]
is fitted, where $q_\mathrm{pre}$ and $q_\mathrm{inst}$ are the coefficients describing the linear dependence of $-\mathrm{ln}(b)$ on the corresponding $x$ variables. 
While the intercept has no relevance in this context, $\hat{\beta}_\mathrm{pre}=\exp(\hat{q}_\mathrm{pre})$ and $\hat{\beta}_\mathrm{inst}=\exp(\hat{q}_\mathrm{inst})$ can be interpreted as the average slopes between tests with pre-analytical treatment B vs.\ A and instrument platform Y vs.\ X, respectively. 
For this specific example, $\hat{\beta}_\mathrm{pre}=0.965$ and $\hat{\beta}_\mathrm{inst}=0.978$
indicate both a difference in slopes due to pre-analytical treatment and a smaller difference due to the instrument platform. 

Similarly, the intercept parameters $a$ (without taking logarithms) may be fitted,
yielding $\Delta \hat{a}_\mathrm{pre}=5.55$ and $\Delta\hat{a}_\mathrm{inst}=3.53$.

This type of analysis requires compatible slope and intercept estimates, which significantly enhances the attractiveness of mPBR compared to pairwise comparisons.

\section{Discussion} 
While the comparison of multiple measurement methods is a common task in clinical chemistry, the current methodology used to analyze these comparisons is limited to repeated pairwise comparisons. This approach becomes inefficient and the resulting reports become more complex as more methods are compared.

In a previous study \cite{Feldmann1981}, a bespoke methodology called mRMR was proposed, but it did not gain much popularity, possibly because even pairwise RMR is not widely used in clinical chemistry.
Therefore, the aim of the present article was to extend the well-established pairwise PBR methodology to the multidimensional setting. The mPBR technique, compared to mRMR, is robust against outlying measurements and also shares the advantages of mRMR, such as compatibility of slope estimators between pairwise measurement comparisons.
The estimators are designed to be equivariant under a scaling of the individual measurements, which allows the approach to be applicable not only to MC, but also to method transformation.
For example, mPBR was applied to the comparison of SARS-CoV-2 serological tests \cite{FERRARI2021144}, some of which were standardized against different standards, resulting in pairwise slopes vastly different from 1.

When comparing $N$ methods using mPBR, only $N-1$ parameters need to be estimated, which allows for the calculation of all $N(N-1)/2$ pairwise slopes and $N-1$ intercepts.
As shown in the example section, this approach leads to a much more economical representation of the parameters (cf.\ Figs.\ \ref{fig:slopesvsintercept}, \ref{fig:Neonatepara}, \ref{fig:Clinpara}) and variance structure (Figs.\ \ref{fig:BlandAltmanMult}, \ref{fig:Neonateba}) compared to pairwise tabulations and Bland--Altman plots. 
The inverse slope and intercept parameters, b and a respectively, also allow for further analyses, such as investigating the factors that determine their values.

As an example, in  section \ref{sec:Clinchem}, the influences of pre-analytical treatment and instrument platform on the parameters obtained for assays from clinical chemistry were analyzed using OLS. 
Especially the possibility to compare several method types (like instrument platform, sample preparation and reagent lot) and whole groups of methods, increases the attractiveness of the regression method considerably.

As far as the mathematical formulation of the mPBR estimator is concerned (cf.\ Sect.\ \ref{sec:StatProp}), it relates to the field of directional statistics \cite{mardia2000directional,fisher1993}, which opens up interesting possibilities for further exploration. 
However, for most of the formal statistical properties of the mPBR slope estimator, such as consistency, efficiency, or finite sample bias, only educated guesses could be formulated in this article. 
Stronger results would require specialization to a specific model setting \cite{cheng1999statistical}. 
Generally, it can be stated that the theoretical basis of MC is by far not as developed as that of ordinary, even nonlinear, regression and brings with it subtle new problems.

In the analysis of MC study data, simply estimating a regression line is not sufficient. 
Confidence intervals for the slope, intercept, and bias at medical decision points should also be reported. 
The most common technique for estimating confidence intervals for regression, including PBR, is bootstrapping \cite{carpenter2000bootstrap}. 
This technique involves repeatedly estimating point estimates on random samples from the original dataset. Bootstrapping can also be applied to mPBR. Once confidence intervals have been determined, questions about the significance of deviations from the slope of 1 or the intercept of 0 can be answered.

Regarding the practical implementation of the proposed algorithm, at the moment, only a basic variant of the Weiszfeld algorithm has been implemented. 
This algorithm scales at least quadratically with the number of observations and requires all samples to have been measured with all methods. 
In the future, it might be possible to impute missing values using an expectation maximization type of algorithm.

\clearpage

\section*{Acknowledgement}
I thank Davide Ferrari for explaining the background of the SARS-CoV-2  study and Jonas Mir for carefully reading the manuscript and very helpful discussions. 

ELECSYS and COBAS  are trademarks of Roche. All other product names and trademarks are the property of their respective owners.

\bibliographystyle{WileyNJD-AMA}
\bibliography{Multidimensional}

\begin{thebibliography}{10}
\providecommand \doibase [0]{http://dx.doi.org/}%

\bibitem{altman1983measurement}
Altman DG, Bland JM. Measurement in medicine: the analysis of method comparison
  studies. {\it J R Stat Soc Ser D} 1983\string; 32(3)\string: 307--317.

\bibitem{bland1986statistical}
Bland JM, Altman D. Statistical methods for assessing agreement between two
  methods of clinical measurement. {\it Lancet} 1986\string; 327(8476)\string:
  307--310.

\bibitem{cheng1999statistical}
Cheng CL, Van~Ness JW. {\it Statistical regression with measurement error}.
\newblock Arnold .
\newblock 1999.

\bibitem{linnet1990estimation}
Linnet K. Estimation of the linear relationship between the measurements of two
  methods with proportional errors. {\it Stat Med} 1990\string; 9(12)\string:
  1463--1473.

\bibitem{linnet1993evaluation}
Linnet K. Evaluation of regression procedures for methods comparison studies.
  {\it Clin Chem} 1993\string; 39(3)\string: 424--432.

\bibitem{francq2014measurement}
Francq BG, Govaerts BB. Measurement methods comparison with errors-in-variables
  regressions. From horizontal to vertical OLS regression, review and new
  perspectives. {\it Chemometr Intell Lab Syst} 2014\string; 134\string:
  123--139.

\bibitem{bolfarine2020regression}
Bolfarine H, De~Castro M, Galea M, others . {\it Regression models for the
  comparison of measurement methods}.
\newblock Springer .
\newblock 2020.

\bibitem{assay2013}
Hojvat S, Kondratovic MV. Assay Migration Studies for In Vitro Diagnostic
  Devices; Guidance for Industry and {FDA} Staff. tech. rep., U.S.\ Department
  of Health and Human Services, Food and Drug Administration, Center for Device
  and Radiological Health; :   2013.

\bibitem{clinical2013ep09}
{Clinical and Laboratory Standards Institute} . {EP09-A3. Measurement procedure
  comparison and bias estimation using patient samples; approved guideline --
  third edition}.  2013.

\bibitem{kummell1879reduction}
Kummell CH. Reduction of observation equations which contain more than one
  observed quantity. {\it The Analyst} 1879\string: 97--105.

\bibitem{adcock1878problem}
Adcock RJ. A problem in least squares. {\it The Analyst} 1878\string;
  5(2)\string: 53--54.

\bibitem{passing1983new}
Passing H, Bablok W. A new biometrical procedure for testing the equality of
  measurements from two different analytical methods. Application of linear
  regression procedures for method comparison studies in clinical chemistry,
  {Part I}. {\it J Clin Chem Clin Biochem} 1983\string; 21\string: 709--720.

\bibitem{bablok1988general}
Bablok W, Passing H, Bender R, Schneider B. A general regression procedure for
  method transformation. {Application} of linear regression procedures for
  method comparison studies in clinical chemistry, {Part III}. {\it J Clin Chem
  Clin Biochem} 1988\string; 26\string: 783--790.

\bibitem{dufey2020derivation}
Dufey F. Derivation of {Passing--Bablok} regression from {Kendall’s} tau.
  {\it Int J Biostat} 2020\string; 16(2)\string: 20190157.

\bibitem{Feldmann1981}
Feldmann U, Schneider B, Klinkers H, Haeckel R. A Multivariate Approach for the
  Biometric Comparison of Analytical Methods in Clinical Chemistry. {\it J Clin
  Chem Clin Biochem} 1981\string; 19(3)\string: 121--138.

\bibitem{fisher1993}
Fisher N, Lunn A, Davies S. Spherical median axes. {\it J R Stat Soc Ser B
  (Method)} 1993\string; 55(1)\string: 117--124.

\bibitem{mardia2000directional}
Mardia KV, Jupp PE, Mardia K. {\it Directional statistics}. 2.
\newblock Wiley Online Library .
\newblock 2000.

\bibitem{FERRARI2021144}
Ferrari D, Clementi N, Span{\`o} SM, et al. Harmonization of six quantitative
  {SARS-CoV-2} serological assays using sera of vaccinated subjects. {\it Clin
  Chim Acta} 2021\string; 522\string: 144--151.

\bibitem{pronzato2013design}
Pronzato L, P{\'a}zman A. {\it Design of experiments in nonlinear models}. 212
  of {\it Lecture notes in statistics}.
\newblock Springer .
\newblock 2013.

\bibitem{weiszfeld1937}
Weiszfeld E. Sur le point pour lequel la somme des distances de n points
  donn{\'e}s est minimum. {\it Tohoku Math J} 1937\string; 43\string: 355--386.

\bibitem{beck2015weiszfeld}
Beck A, Sabach S. {Weiszfeld’s} method: Old and new results. {\it J Optim
  Theory Appl} 2015\string; 164(1)\string: 1--40.

\bibitem{dempster1977maximum}
Dempster AP, Laird NM, Rubin DB. Maximum likelihood from incomplete data via
  the {EM} algorithm. {\it J R Stat Soc: Ser B (Method)} 1977\string;
  39(1)\string: 1--22.

\bibitem{raymaekers2022equivariant}
Raymaekers J, Dufey F. Equivariant {Passing--Bablok} regression in quasilinear
  time. {\it arXiv preprint arXiv:2202.08060} 2022.

\bibitem{reiersol1950identifiability}
Reiers{\o}l O. Identifiability of a linear relation between variables which are
  subject to error. {\it Econometrica} 1950\string; 18\string: 375--389.

\bibitem{thomas2022total}
Thomas DH, Warner JV, Jones GR, et al. Total bilirubin assay differences may
  cause inconsistent treatment decisions in neonatal hyperbilirubinaemia. {\it
  Clin Chem Lab Med} 2022\string; 60(11)\string: 1736--1744.

\bibitem{carpenter2000bootstrap}
Carpenter J, Bithell J. Bootstrap confidence intervals: when, which, what? A
  practical guide for medical statisticians. {\it Stat Med} 2000\string;
  19(9)\string: 1141--1164.

\end{thebibliography}

\end{document}